\documentclass[preprint,eqsecnum,aps,showpacs,nofootinbib]{revtex4}
\pdfoutput=1
\usepackage{bm,amssymb,amsmath,cancel,footnote,subfigure}
 \usepackage{feynmp}
 \usepackage[pdftex]{graphicx}
\begin{document}

\count255=\time\divide\count255 by 60 \xdef\hourmin{\number\count255}
  \multiply\count255 by-60\advance\count255 by\time
 \xdef\hourmin{\hourmin:\ifnum\count255<10 0\fi\the\count255}

\newcommand{\xbf}[1]{\mbox{\boldmath $ #1 $}}
\newcommand{\Dslash}{D\hspace{-0.7em}{ }\slash\hspace{0.2em}}

\title{Precision Electroweak Constraints on the $N=3$ Lee-Wick Standard
Model}

\author{Richard F. Lebed}
\email{Richard.Lebed@asu.edu}

\author{Russell H. TerBeek}
\email{r.terbeek@asu.edu}

\affiliation{Department of Physics, Arizona State University, 
Tempe, AZ 85287-1504}


\date{October 2012}

\begin{abstract}
  The Lee-Wick (LW) formulation of higher-derivative theories can be
  extended from one in which the extra degrees of freedom are
  represented as a single heavy, negative-norm partner for each known
  particle ($N=2$), to one in which a second, positive-norm partner
  appears ($N=3$).  We explore the extent to which the presence of
  these additional states in a LW Standard Model affect precision
  electroweak observables, and find that they tend either to have a
  marginal effect ({\it e.g.}, quark partners on $T$), or a
  substantial beneficial effect ({\it e.g.}, Higgs partners on the
  $Zb\bar b$ couplings).  We find that precision constraints allow LW
  partners to exist in broad regions of mass parameter space
  accessible at the LHC, making LW theories a viable beyond-Standard
  Model candidate.
\end{abstract}


\maketitle
\thispagestyle{empty}

\newpage
\setcounter{page}{1}

\section{Introduction} \label{sec:intro}

If the particle of mass $126$~GeV recently
discovered~\cite{:2012gk,:2012gu} at the Large Hadron Collider (LHC)
turns out (as is widely expected) to be the Higgs scalar, then
particle physics will have at last undeniably moved into the
beyond-Standard Model (BSM) era.  The theoretical difficulties of a
universe in which the Standard Model (SM) is the ultimate theory of
particle physics are well known: In addition to requiring three
complete generations of fermions, and ignoring gravity but
nevertheless incorporating three distinct fundamental interactions,
the SM suffers from the famous hierarchy problem of a scalar particle
whose renormalized mass lies quite close to the scale of electroweak
symmetry breaking, rather than being driven to GUT- or Planck-scale
values by the exigencies of regularizing a quadratic divergence.  The
most popular BSM remedies for the hierarchy problem are also well
known: Low-scale supersymmetry (SUSY), large extra spacetime
dimensions, and little Higgs models.  As the LHC continues to generate
vast amounts of new experimental data, the constraints of
phenomenological viability are pushing each approach into ever smaller
regions of its respective parameter space.  The moment of truth for
many BSM models is rapidly approaching.

The same can be said for a less well-studied approach, the Lee-Wick
Standard Model (LWSM) of Grinstein, O'Connell, and
Wise~\cite{Grinstein:2007mp}.  Inspired by the Lee and Wick (LW)
program~\cite{Lee:1970iw} of performing renormalization by promoting
the spurious Pauli-Villars regulator to the status of a full,
dynamical, negative-norm field, Ref.~\cite{Grinstein:2007mp} showed
that introducing LW partners for SM particles with the same gauge
couplings eliminates quadratic divergences in loop calculations.  The
cancellation between positive- and negative-norm states in loops
resembles the cancellation between fermions and bosons in SUSY, while
the fact that the particle and its LW partner share the same
statistics but carry an opposite type of parity is reminiscent of the
bottom of a tower of Kaluza-Klein excitations in extra-dimension
models.

The latter analogy becomes more apparent when one realizes that LW
models need not terminate with a single partner.  As shown in
Ref.~\cite{Grinstein:2007mp}, the LW Lagrangian is equivalent to a
particular higher-derivative (HD) theory; in particular, it is one in
which 4-derivative bosonic and 3-derivative fermionic interaction
terms appear, and the full HD field consists of both the conventional
field and its LW partner.  Of course, not just any HD Lagrangian
produces an equivalent LW theory; only those that produce propagator
poles at real mass values are valid for the purpose.  Labeling
theories by $N$, the number of poles in the HD field propagator, the
conventional single-pole theory is labeled as $N=1$, and the original
LW theory is labeled as $N=2$, but in principle nothing prevents the
construction of $N \ge 3$ theories~\cite{Carone:2008iw}.  In such
theories, one can show that the partner states alternate in norm as
their mass parameters increase.  The cancellation of quadratic
divergences requires the participation of all $N$ states through
delicate sum rules among their couplings that seem conspiratorial at
the level of the LW theory, but merely reflect the improved power
counting of the equivalent HD theory.

While not as thoroughly studied as other BSM approaches, the original
LWSM approach~\cite{Grinstein:2007mp} has nevertheless inspired
research leading to numerous publications in several different areas,
including early universe models, quantum gravity, thermodynamics, and
formal studies of field theory.  The last of these deserves special
mention because negative-norm states in field theory are peculiar
objects.  As has been known for decades~\cite{Coleman:1969xz}, the
apparent violation of unitarity induced by such states can be traded
for the imposition of future boundary conditions that introduce
causality violation at microscopic levels.  To date, no logical
argument precludes the existence of such exotic behavior, and the
existence of microcausality violation can only be bounded
experimentally by measurements at successively higher energy scales.

For the purposes of this paper, we avoid such thorny issues and adopt
instead the pragmatic viewpoint that LW theories (or their HD
equivalents) should merely be treated as effective theories good to
scales of at least 14~TeV, the upper limit of physics to be probed at
the LHC in the near future.  The question of the viability of LWSM
variants then relies upon whether the new states can be produced and
observed directly, and for what mass ranges they satisfy the stringent
experimental constraints imposed by electroweak precision tests
(EWPT).  Both of these questions have been studied in some detail in
the original $N=2$ LWSM; in the case of direct production,
Refs.~\cite{Rizzo:2007ae,Rizzo:2007nf} find that $N=2$ LW gauge
bosons, for example, can readily be produced at the LHC, but may be
difficult to distinguish from novel states from other scenarios such
as extra-dimension models.  Precision observables in the $N=2$ theory,
on the other hand, have been examined in a succession of
improvements~\cite{Alvarez:2008za,Underwood:2008cr,Carone:2008bs,
Chivukula:2010nw} (by scanning the LW parameter space
in~\cite{Alvarez:2008za}; by including only LW masses for the fields
most important for the hierarchy problem~\cite{Carone:2008bs}; by
using not just oblique parameters $S$, $T$, but also the ``post-LEP''
parameters $W$, $Y$~\cite{Underwood:2008cr}; by including bounds from
the $Z b \bar b$ direct correction~\cite{Chivukula:2010nw}), with the
consensus conclusion that LW gauge boson masses must be well over
$2$~TeV, and in such cases, the LW fermion masses must be
substantially higher (perhaps as much as 10~TeV).  If all LW masses
are comparable, then the lower bound on this scale is typically $\sim
7$~TeV\@.  The LW Higgs partners, on the other hand, appear to be much
less tightly constrained and produce milder constraints on collider
phenomenology~\cite{Krauss:2007bz,Carone:2009nu,Alvarez:2011ah,
Figy:2011yu}.

In comparison, only one collider physics study of the $N=3$ LWSM has
thus far appeared~\cite{Lebed:2012zv}, a paper by the present authors
generalizing the study of $W$ boson production in
Ref.~\cite{Rizzo:2007ae}, and showing not only that such bosons can
readily be produced, but also that their mass spectrum generates a
signature likely unique among known BSM models.  The next logical step
is, of course, a study of EWPT in the $N=3$ LWSM, which is the purpose
of this paper.

On general principles, one naturally expects the $N=3$ LWSM to allow
for less stringent lower bounds on new particle masses compared to the
$N=2$ model, making for earlier discovery potential at the LHC\@.  Of
course, simply by adding new degrees of freedom to the theory
(extending from $N=2$ to $N=3$) and then fitting to EWPT, one expects
the bounds to relax; however, in LW models, one might expect the
effect to be more pronounced because the negative-norm states and the
new positive-norm states can produce a substantial numerical
cancellation just between themselves (although the SM state must also
be included in order to cancel the quadratic divergences).  Since the
$N=2$ LWSM may be thought of as an $N=3$ model in which the masses of
the negative-norm states are fixed and the masses of the additional
positive-norm states are taken to infinity, one expects a substantial
relaxation of tension in EWPT compared to the $N=2$ LWSM when the
positive-norm masses are adjusted to lie not excessively higher than
the negative-norm masses.  In detailed fits, we find that this
reasoning holds up to scrutiny in the scalar sector, while the
addition of $N=3$ fermions generates much more nuanced changes,
sometimes even moving in the same direction as the $N=2$ contribution.
After a detailed analysis, one finds that a large parameter space of
LHC-accessible masses remains open to LW partner states, making the
$N=3$ LWSM phenomenologically viable and attractive.

This paper is organized as follows.  In Sec.~\ref{sec:N=3} we review
the formalism of the $N=3$ LWSM\@.  Section~\ref{sec:oblique} defines
the oblique EWPT parameters used in the fits, while Sec.~\ref{sec:dg}
considers an important non-oblique EWPT variable, the $Z b_L \bar b_L$
coupling.  In Sec.~\ref{sec:analysis} we analyze the effects of EWPT
and present bounds on the $N=3$ LWSM particle masses.
Section~\ref{sec:concl} offers discussion and concluding remarks.

\section{Review of the $N=3$ Lee-Wick Standard Model} \label{sec:N=3}

A Lee-Wick theory of degree $N$ for a given field $\hat \phi$ is a
particular higher-derivative theory in which the original Lagrangian
with a canonical kinetic energy term is augmented by the addition of
terms containing up to $2N$ additional covariant derivatives.  Such a
Lagrangian may be re-expressed in terms of an equivalent auxiliary
field formalism in which $\hat \phi$ is a linear combination of $N$
fields $\phi^{(1),(2),\ldots,(N)}$ that alternate in the sign of their
quantum-mechanical norm.  As shown in Ref.~\cite{Carone:2008iw} and
summarized in this section, this construction can be implemented
independently for fields $\hat \phi$ that are real or complex scalars,
fermions, or gauge fields.  In particular, no obvious theory
constraint fixes the mass parameters that appear with each additional
pair of derivatives acting upon each field, so that one may consider
scenarios, for example, in which only some of the SM particles have
one LW partner, some have two, and some have none.

In the $N=2$ LW theory, the opposite-sign norms are incorporated by
the fields corresponding to particles and their partners that appear
in the Lagrangian with a relative sign, {\it i.e.}, $\hat \phi =
\phi^{(1)} - \phi^{(2)}$.  For any integer $N > 2$, the origin of the
equivalence between the LW theory and its HD form is imposed by means
of a set of fixed parameters $\eta_{1,2,\ldots,N}$.  For $N = 3$ they
read~\cite{Carone:2008iw}
\begin{eqnarray}
\eta_1 &\equiv & \frac{\Lambda^4}{(m_2^2-m_1^2)(m_3^2-m_1^2)} \, ,
\label{eq:etadef1} \\
\eta_2 &\equiv & \frac{\Lambda^4}{(m_1^2-m_2^2)(m_3^2-m_2^2)} \, ,
\label{eq:etadef2}\\
\eta_3 &\equiv & \frac{\Lambda^4}{(m_1^2-m_3^2)(m_2^2-m_3^2)} \, ,
\label{eq:etadef3}
\end{eqnarray}
where $m_1 < m_2 < m_3$ are the masses of the original state and its
two LW partners, and $\Lambda^4 \equiv m_1^2 m_2^2 + m_1^2 m_3^2 +
m_2^2 m_3^2$.  The parameters satisfy a variety of sum rules,
\begin{equation} 
\sum_{i=1}^3 m_i^{2n} \,\eta_i =0  \ \ (n=0,1),
\label{eq:sr1}
\end{equation}
\begin{equation}
\sum_{i=1}^3 m_i^{2n} \,\eta_i =\Lambda^4  \ \ (n=2),
\label{eq:sr2}
\end{equation}
\begin{equation}
m_1^2 m_2^2 \eta_3 + m_2^2 m_3^2 \eta_1 + m_3^2 m_1^2 \eta_2 =
\Lambda^4 \, .
\label{eq:sr3}
\end{equation}
that provide the means by which cancellations of quadratic loop
divergences are guaranteed.  They appear in slightly different
permutations in fields of different spin.

\subsection{Neutral Scalar Fields}
Upon writing
\begin{equation}
\hat \phi =  \sqrt{\eta_1} \, \phi^{(1)} \! - \sqrt{-\eta_2} \,
\phi^{(2)} + \sqrt{\eta_3} \, \phi^{(3)}  \, , \label{eq:redef1} \\
\end{equation}
an $N=3$ HD Lagrangian of the general form
\begin{equation}
{\cal L}_{{\rm HD}}^{N=3}=-\frac{1}{2} \hat{\phi} \,\Box\, \hat{\phi}
- \frac{1}{2M_1^2} \hat{\phi} \,\Box^2 \hat{\phi}- \frac{1}{2M_2^4}
\hat{\phi} \,\Box^3 \hat{\phi} -\frac{1}{2} m_\phi^2 \hat{\phi}^2
+{\cal L}_{{\rm int}}(\hat{\phi})
\label{eq:toyhd3}
\end{equation}
is equivalent at the quantum level to the LW Lagrangian (note the
alternation of norm):
\begin{equation}
{\cal L}_{\rm LW}^{N=3} = - \frac{1}{2} \phi^{(1)}
( \Box \, + \, m_1^2 ) \,
\phi^{(1)} + \frac{1}{2} \phi^{(2)} ( \Box \, + \, m_2^2 ) \,
\phi^{(2)} - \frac{1}{2} \phi^{(3)} ( \Box \, + \, m_3^2 ) \,
\phi^{(3)} + {\cal L}_{{\rm int}}( \hat{\phi} ) \, ,
\label{eq:m10res}
\end{equation}
provided one identifies
\begin{eqnarray}
m_\phi^2 & = & (m_1^2 m_2^2 m_3^2)/\Lambda^4 \, ,
\label{eq:msqds1} \\
M_1^2 & = & \Lambda^4/(m_1^2+m_2^2+m_3^2) \, ,
\label{eq:msqds2} \\
M_2^2 & = & \Lambda^2 \, . \label{eq:msqds3}
\end{eqnarray}

\subsection{Yang-Mills Fields}
The analogue to Eq.~(\ref{eq:redef1}) reads
\begin{equation}
\hat A^\mu = A_1^\mu - \sqrt{\frac{-\eta_2}{\ \ \eta_1}} A_2^\mu
+ \sqrt{\frac{\eta_3}{\eta_1}} A_3^\mu \, ,
\end{equation}
with $m_1$ set to zero to guarantee the masslessness of the gauge
field $A_1^\mu$.  One defines the field strength and covariant
derivative acting upon an adjoint representation field $X$ in the
usual way:
\begin{eqnarray}
\hat F^{\mu \nu} & \equiv & \partial^\mu \hat A^\nu - \partial^\nu
\hat A^\mu - ig \,[ \hat A^\mu , \hat A^\nu ] \, , \label{fs} \\
\hat D^\mu X & \equiv & \partial^\mu X - ig \,[ \hat A^\mu , X ] \, .
\label{cov}
\end{eqnarray}
Then the $N=3$ HD Lagrangian,
\begin{equation}
{\cal L}_{\rm HD}^{N=3} = -\frac 1 2 \, {\rm Tr} \, \hat F_{\mu \nu}
\hat F^{\mu \nu} - \left( \frac{1}{m_2^2} \! + \! \frac{1}{m_3^2}
\right)
{\rm Tr} \hat F_{\mu \nu} \hat D^\mu \hat D_\alpha \hat F^{\alpha \nu}
- \frac{1}{m_2^2 m_3^2} \, {\rm Tr} \hat F_{\mu \nu} \hat D^\mu \hat
D_\alpha \hat D^{[\alpha} \hat D_\beta \hat F^{\beta \nu ]} \, ,
\label{YMHD}
\end{equation}
where the superscript brackets indicate antisymmetrization of just the
first and last indices ($\alpha$ and $\nu$ here), is equivalent to the
LW Lagrangian
\begin{eqnarray}
{\cal L}_{\rm LW}^{N=3} & = & -\frac 1 2 \, {\rm Tr} \, F_1^{\mu \nu}
F_{1 \mu \nu}
+ \frac 1 2 \, {\rm Tr} (D_\mu A_{2 \nu} - D_\nu A_{2 \mu} )^2 - \frac
1 2 \, {\rm Tr} ( D_\mu A_{3 \nu} - D_\nu A_{3 \mu} )^2
\nonumber \\ & & - m_2^2 \, {\rm Tr} A_2^\mu A_{2 \mu} + m_3^2
\, {\rm Tr} A_3^\mu A_{3 \mu} \, ,
\end{eqnarray}
which includes all of the kinetic and mass terms, plus more involved
but still fairly compact expressions for cubic and quartic terms given
explicitly in Ref.~\cite{Carone:2008iw}.  The alternation of norm is
again apparent.

\subsection{Chiral Fermion Fields}
Chiral fermions are only slightly more complicated because their LW
partners have explicit LW Dirac mass partners.  For a conventional
left-handed Weyl fermion field $\phi_L$, the analogue of
Eq.~(\ref{eq:redef1}) reads
\begin{equation}
\hat \phi_L = \phi^{(1)}_L - \sqrt{\frac{-\eta_2}{\ \ \eta_1}} \,
\phi^{(2)}_L + \sqrt{\frac{\eta_3}{\eta_1}} \, \phi^{(3)}_L \, ,
\label{HDfermion}
\end{equation}
and the LW partner fields $\phi^{(2),(3)}_L$ possess their own chiral
partners $\phi^{(2),(3)}_R$ that arise from the process of converting
the HD Lagrangian into an equivalent LW form.  Defining then for each
LW partner the combined field $\phi \equiv \phi_L + \phi_R$ and noting
that $m_1 = 0$, the HD form reads
\begin{equation}
{\cal L}_{\rm HD}^{N=3} = \frac{1}{m_2^2 m_3^2} \overline{\hat \phi}_L
\left [ ( i \hat{\Dslash} )^2 - m_2^2 \right] \left [ ( i \hat{\Dslash}
)^2 - m_3^2 \right] i \hat{\Dslash} \hat \phi_L \ , \label{HDf}
\end{equation}
where $\hat\Dslash$ includes both the gauge bosons and their LW
partners.  The equivalent LW Lagrangian then reads
\begin{equation}
{\cal L}_{\rm LW}^{N=3} = \overline{\phi}^{(1)}_L i \hat{\Dslash}
\phi^{(1)}_L  - \overline{\phi}^{(2)} \! ( i \hat{\Dslash} - m_2 )
\phi^{(2)}    + \overline{\phi}^{(3)} \! ( i \hat{\Dslash} - m_3 )
\phi^{(3)} \, .
\end{equation}
In the case of a fundamental right-handed Weyl field $\phi_R$
contained in a HD Lagrangian field $\hat \phi_R$, the definitions
proceed exactly as above, with the substitution $L \leftrightarrow R$.
However, one should note that the $R$ chiral partners induced in the
$\hat \phi_L$ construction are distinct fields from those appearing
directly in the definition $\hat \phi_R$, and vice versa for $L$
chiral partners.

The original paper~\cite{Grinstein:2007mp} adopts the notation of
placing a prime on fields that appear not through HD superfields but
rather through their Dirac mass terms\footnote{In contrast,
Ref.~\cite{Chivukula:2010nw} uses primes exclusively for the
right-handed HD superfields and Dirac mass partners of its component
fields.}; for example, in the third generation, the SM fields $t_L$,
$b_L$ transforming under SU(2)$\times$U(1) as ({\bf 2}, $+\frac 1 6$)
are joined by $N=2$ LW partners $\tilde t_L$, $\tilde b_L$, and the
latter have Dirac mass partners (mass parameter $M_q$) $\tilde
t^\prime_R$, $\tilde b^\prime_R$, respectively, all of which transform
as ({\bf 2}, $+\frac 1 6$).  The SM fields $t_R$ and $b_R$,
transforming as ({\bf 1}, $+\frac 2 3$) and ({\bf 1}, $-\frac 1 3$),
respectively, have $N=2$ LW partners $\tilde t_R$, $\tilde b_R$, which
in turn have Dirac mass partners $\tilde t^\prime_L$ (mass $M_t$),
$\tilde b^\prime_L$ (mass $M_b$), respectively.  For $N > 2$, we
retain the prime convention of~\cite{Grinstein:2007mp}, replace the
tildes with superscripts $(2),(3),\ldots ,$ and attach corresponding
subscripts to the masses ({\it e.g.}, $M_{q2}$, $M_{b3}$).  For
purposes of numerical analysis, the fields are more conveniently
collected~\cite{Carone:2008bs} by flavor and chirality, rather than by
SU(2)$\times$U(1) quantum numbers.  In the $N=3$ case,
\begin{eqnarray}
  T_{L,R}^T & \equiv & \left( t_{L,R}^{(1)}, \, t_{L,R}^{(2)}, \,
t_{L,R}^{' \, (2)}, \, t_{L,R}^{(3)}, \, t_{L,R}^{' \, (3)} \right)
\, , \nonumber \\
  B_{L,R}^T & \equiv & \left( b_{L,R}^{(1)}, \, b_{L,R}^{(2)}, \,
b_{L,R}^{' \, (2)}, \, b_{L,R}^{(3)}, \, b_{L,R}^{' \, (3)} \right)
\, . \label{superfields}
\end{eqnarray} 

\subsection{Complex Scalar Fields} \label{complex}
The generalization of the real scalar field $\phi$ to a complex scalar
multiplet $H$ transforming in the fundamental representation of a
non-Abelian gauge group requires only the promotion of ordinary
derivatives to covariant ones.  The analogue of Eq.~(\ref{eq:redef1})
reads
\begin{equation} \label{scalarweights}
\hat{H} = \sqrt{\eta_1} H^{(1)} - \sqrt{-\eta_2} H^{(2)} +
\sqrt{\eta_3} H^{(3)} \, ,
\end{equation}
and relates the HD form,
\begin{equation}
{\cal L}_{\rm HD}^{N=3} = \hat D_\mu \hat{H}^\dagger \hat D^\mu
\hat{H} -m_H^2 \hat{H}^\dagger \hat{H}- \frac{1}{M_1^2} 
\hat{H}^\dagger (\hat D_\mu \hat D^\mu)^2 \hat{H} - \frac{1}{M_2^4}
\hat{H}^\dagger (\hat D_\mu \hat D^\mu)^3 \hat{H}  +
{\cal L}_{{\rm int}}(\hat{H}) \, ,
\end{equation}
to the equivalent LW form
\begin{eqnarray}
{\cal L}_{\rm LW}^{N=3} &=&
- H^{(1)\dagger}(\hat D_\mu \hat D^\mu + m_1^2) H^{(1)}
+ H^{(2)\dagger} ( \hat D_\mu \hat D^\mu+m_2^2) H^{(2)}
- H^{(3)\dagger} ( \hat D_\mu \hat D^\mu+m_3^2) H^{(3)}
\nonumber \\ && + {\cal L}_{{\rm int}} (\hat{H})  \, ,
\end{eqnarray}
with the mass parameters related as in
Eqs.~(\ref{eq:msqds1})--(\ref{eq:msqds3}), with $m_\phi \to m_H$.

In the particular case of the SM Higgs multiplet, $m_1 = 0$, and the
lightest scalar obtains mass only through spontaneous symmetry
breaking with vacuum expectation value $v$.  Writing
\begin{eqnarray}
{\cal L}_{\rm HD}^{N=3} & = & {\cal L}_{\rm HD}^{N=3} (m_H^2 = 0) +
\tilde {\cal L}_{{\rm int}}(\hat{H}) \, , \\
-\tilde {\cal L}_{\rm int}(\hat{H}) & \equiv & \frac{\lambda}{4}
\left( \hat{H}^\dagger \hat{H}  - \frac{v^2}{2} \right)^2 \, ,
\end{eqnarray}
the equivalent LW Lagrangian reads
\begin{eqnarray}
{\cal L}_{\rm LW}^{N=3} &=& \hat D_\mu H^{(1)\dagger} \hat D^\mu
H^{(1)} - \hat D_\mu H^{(2)\dagger} \hat D^\mu H^{(2)} +
\hat D_\mu H^{(3)\dagger} \hat D^\mu H^{(3)} \nonumber \\ && +
m_2^2 H^{(2)\dagger} H^{(2)}
-m_3^2 H^{(3)\dagger} H^{(3)} + \tilde {\cal L}_{{\rm int}}
( \hat H ) \, .
\label{eq:llp}
\end{eqnarray}
In unitary gauge,
\begin{equation} \label{Higgses}
H^{(1)} = \left(\begin{array}{c} 0 \\ \frac{1}{\sqrt{2}}(v+h_1)
\end{array} \right) , \,\,\,\,\,
H^{(2)} = \left(\begin{array}{c} ih_2^+ \\ \frac{1}{\sqrt{2}}
(h_2 + i P_2) \end{array} \right) ,
\,\,\,\,\,
H^{(3)} = \left(\begin{array}{c} ih_3^+ \\ \frac{1}{\sqrt{2}}
(h_3 + i P_3) \end{array} \right) ,
\end{equation}
where the fields $h_i$, $P_i$, and $h^+_i$ denote the scalar,
pseudoscalar, and charged Higgs components, respectively, the mass
terms in Eq.~(\ref{eq:llp}) read
\begin{eqnarray}
{\cal L}_{\rm mass}^{N=3} &=& \frac{1}{2} m_2^2 \,
(2 h_2^- h_2^+ + h_2^2 + P_2^2)
-\frac{1}{2} m_3^2 \, (2 h_3^- h_3^+ + h_3^2 + P_3^2) \nonumber \\ 
&&- \frac{1}{2} m^2 (h_1 - \sqrt{-\eta_2} h_2 +\sqrt{\eta_3} h_3)^2
\, , \label{scalarmasses}
\end{eqnarray}
with $m^2 = \lambda v^2/2$.  The pseudoscalar and charged scalar
fields therefore have mass eigenvalues $m_{2,3}$, while the neutral
scalar fields are mixed.  The mass eigenvectors $h^0$ in the mixed
sector are obtained by a symplectic transformation $S$ that preserves
the relative signs of the kinetic terms via a metric $\eta = {\rm
diag} ( +, -, + )$ but diagonalizes the mass matrix ${\cal M}$ in
$h^\dagger {\cal M} \eta h$:
\begin{equation}
h^0 = S^{-1} h \, , \hspace{2em} S^\dagger \eta S = \eta \, ,
\end{equation}
so that
\begin{equation}
{\cal M}_0 \eta = S^\dagger {\cal M} \eta S \, .
\end{equation}
In the $N=2$ case~\cite{Grinstein:2007mp}, the elements of $S$ consist
of $\sinh \phi$ and $\cosh \phi$ of a single ``Euler angle'' $\phi$.
For higher $N$, $S$ is similarly expressible as the symplectic
analogue to a multidimensional Euler rotation matrix.  In any case,
the transformation $S$ for any given mixing matrix ${\cal M}$ is
easily found numerically.

\subsection{Fermion Mass Diagonalization}
Since the Yukawa couplings appear as
\begin{equation}
{\cal L}_{\rm Yuk} = - y_t \, \, \hat \bar \hspace{-0.5ex} q_L \hat H
\hat b_R - y_b \, \, \hat \bar \hspace{-0.5ex} q_L (\epsilon
\hat H^\dagger ) \, \hat t_R + {\rm H.c.} \, ,
\end{equation}
where $\epsilon \equiv i\sigma_2$, the fermion mass terms may be
expressed in terms of the ratios of $\eta$'s appearing in
Eq.~(\ref{HDfermion}).  In the case of $t$ quarks for $N=3$, one may
abbreviate $m_t \equiv y_t v / \sqrt{2}$ and:
\begin{eqnarray}
\cosh \phi_q = \frac{M_{q3}}{\sqrt{M_{q3}^2 - M_{q2}^2}} \, , &
\hspace{3em} &
\sinh \phi_q = \frac{M_{q2}}{\sqrt{M_{q3}^2 - M_{q2}^2}} \, ,
\nonumber \\
\cosh \phi_t = \frac{M_{t3}}{\sqrt{M_{t3}^2 - M_{t2}^2}} \, , &
\hspace{3em} &
\sinh \phi_t = \frac{M_{t2}}{\sqrt{M_{t3}^2 - M_{t2}^2}} \, ,
\end{eqnarray}
which give mass terms, using the notation of Eq.~(\ref{superfields}),
of the form
\begin{equation}
{\cal L}^{\rm N=3}_{t \, {\rm mass}} =
- \overline T_L \eta {\cal M}^\dagger_t T_R + {\rm H.c.} \, ,
\end{equation}
where
\begin{equation} \label{massmatrix}
{\cal M}^{N=3}_t \eta = \left( \begin{array}{ccccc}
m_t & -m_t \cosh \phi_q & 0 & m_t \sinh \phi_q & 0 \\
-m_t \cosh \phi_t & m_t \cosh \phi_q \, \cosh \phi_t & -M_{t2} & -m_t
\sinh \phi_q \, \cosh \phi_t & 0 \\
0 & -M_{q2} & 0 & 0 & 0 \\
m_t \sinh \phi_t & -m_t \cosh \phi_q \, \sinh \phi_t & 0 & m_t \sinh
\phi_q \, \sinh \phi_t & +M_{t3} \\
0 & 0 & 0 & +M_{q3} & 0
\end{array} \right) \, ,
\end{equation}
where the metric $\eta = {\rm diag}(+,-,-,+,+)$ reflects the norms of
the component states, and thus also appears in the corresponding
kinetic terms.  The diagonalization of the mass matrix to a form
${\cal M}_{t0}$ with positive eigenvalues therefore requires
independent transformation matrices $S_{L,R}^t$ for each quark flavor
(here, $t$) satisfying the constraints
\begin{equation} \label{massxfm}
S^\dagger_L \eta S_L = \eta \, , \ \ S^\dagger_R \eta S_R = \eta \, ,
\ \ {\cal M}_0 \eta = S^\dagger_R {\cal M} \eta S_L \, ,
\end{equation}
so that the mass eigenstates are obtained as
\begin{equation} \label{masseigen}
T_{L,R}^0 = (S_{L,R}^t)^{-1} T_{L,R} \, ,
\end{equation}
and similarly for the $B$ sector.  Obtaining numerical solutions for
$S_{L,R}^t$ is most efficiently accomplished by converting this system
into an equivalent eigenvalue problem~\cite{Figy:2011yu}.

\section{Bounds on Oblique Parameters} \label{sec:oblique}

\subsection{Formalism and Tree-Level Contributions}

Bounds on BSM physics are typically expressed in terms of oblique
(flavor-universal, arising from gauge boson vacuum polarization loops)
and direct (flavor-specific, arising from vertex, box, {\it etc.},
corrections) parameters~\cite{Kennedy:1988sn}.  The best-known
oblique electroweak observables are the dimensionless Peskin-Takeuchi
(PT) parameters~\cite{PandT} $S$, $T$, $U$, which represent all
independent finite combinations obtained from differences of the
vacuum polarization functions and their first derivatives.  As better
data (particularly from LEP2) became available in the 1990s, probing
the oblique corrections to second-derivative order became possible;
Barbieri {\it et al.}~\cite{Barbieri:2004qk} developed a complete set
of such ``post-LEP'' parameters, $\hat S$, $\hat T$, $\hat U$ (the PT
parameters with different normalizations\footnote{Note that the vacuum
  polarization functions $\Pi (q^2)$ of Ref.~\cite{Barbieri:2004qk}
  are opposite in sign to those as defined in Ref.~\cite{PandT}.}),
$V$, $W$, $X$, $Y$, and $Z$.  Just as Ref.~\cite{PandT} argued that
$U$ is numerically small, Ref.~\cite{Barbieri:2004qk} argued that $V$,
$X$, and $Z$ can be neglected in EWPT, leaving only $\hat S$, $\hat
T$, $W$, and $Y$ as the important independent oblique parameters.

The primitive electroweak parameters are obtained in
Ref.~\cite{Barbieri:2004qk} as:
\begin{eqnarray}
\frac{1}{g^{\prime 2}} \equiv \Pi^\prime_{\hat{B} \hat{B}}(0) \, ,
\ \ 
\frac{1}{g^2} \equiv \Pi^\prime_{\hat{W}^+ \hat{W}^-}(0) \, ,
\label{eq:ggprime} \\
\frac{1}{\sqrt{2}G_F} = - 4 \Pi_{\hat{W}^+ \hat{W}^-}(0) = v^2 \ .
\label{eq:GF}
\end{eqnarray}
In the tree-level SM, these just give the usual parameters $g'= g_1$,
$g = g_2$ and $v$; however, these relations persist in the LWSM as
well.  The reciprocal powers of coupling constant arise from the
choice of a noncanonical normalization of the field
strengths~\cite{Barbieri:2004qk} designed to give a convenient
separation of $g'$, $g$, and $v$ in
Eqs.~(\ref{eq:ggprime})--(\ref{eq:GF}).  From Eq.~(\ref{YMHD}) one
quickly extracts for the $N = 3$ model [where, {\it e.g.}, $M_1^{(3)}$
indicates the 3$^{\rm rd}$ LW partner mass for the U(1) SM gauge
group]:
\begin{eqnarray}
\Pi_{\hat{W}^+ \hat{W}^-}(q^2) = \Pi_{\hat{W}^3 \hat{W}^3}(q^2) &=&
\frac{q^2}{g_2^2} -\frac{(q^2)^2}{g_2^2} \left[
\frac{1}{M_2^{(2) \, 2}} + \frac{1}{M_2^{(3) \, 2}} \right] -
\frac{v^2}{4} \, , \nonumber \\
\Pi_{\hat{W}^3 \hat{B}}(q^2) &=& \frac{v^2}{4} \, , \nonumber \\
\Pi_{\hat{B} \hat{B}}(q^2) &=& \frac{q^2}{g_1^2}
-\frac{(q^2)^2}{g_1^2} \left[ \frac{1}{M_1^{(2) \, 2}} +
\frac{1}{M_1^{(3) \, 2}} \right] -\frac{v^2}{4} \, ,
\label{eq:twopoint}
\end{eqnarray}
from which one sees that the relations $g^\prime = g_1$, $g = g_2$,
and Eq.~(\ref{eq:GF}) are preserved.  In addition, one can easily
compute the tree-level oblique electroweak parameters as done for the
$N = 2$ model in Ref.~\cite{Underwood:2008cr}:
\begin{eqnarray}
\hat{S} &\equiv & g^2\ \Pi^\prime_{\hat{W}^3 \hat{B}}(0) = 0 \, ,
\label{eq:treelevelShat} \\
\hat{T} &\equiv & \frac{g^2}{m_W^2} \left[\Pi_{\hat{W}^3
\hat{W}^3}(0)- \Pi_{\hat{W}^+ \hat{W}^-}(0)\right] = 0 \ ,
\label{eq:treelevelThat} \\
W & \equiv & \frac{1}{2}g^2m_W^2 \
\Pi^{\prime\prime}_{\hat{W}^3 \hat{W}^3}(0)= -m_W^2 \left[
  \frac{1}{M_2^{(1) \, 2}} + \frac{1}{M_2^{(2) \, 2}} \right] \, ,
\label{eq:treelevelW} \\
Y & \equiv & \frac{1}{2} g^{\prime \, 2} m_W^2 \
\Pi^{\prime\prime}_{\hat{B} \hat{B}}(0)= -m_W^2 \left[
  \frac{1}{M_1^{(2) \, 2}} + \frac{1}{M_1^{(3) \, 2}} \right] \, ,
\label{eq:treelevelY}
\end{eqnarray}
Here, the first equality in each equation defines the corresponding
post-LEP parameter~\cite{Barbieri:2004qk}.  The absence of tree-level
contributions to $\hat S$ and $\hat T$ was first noted in
Ref.~\cite{Underwood:2008cr}.  Moreover, Ref.~\cite{Chivukula:2010nw}
noted that the scheme defining Eq.~(\ref{eq:treelevelY}) precludes
fermionic one-loop corrections to $Y$, while $W$ (which is defined in
terms of $\Pi_{\hat{W}^3 \hat{W}^3}$ rather than $\Pi_{\hat{W}^+
  \hat{W}^-}$) was found to have fermionic one-loop corrections that
are numerically small compared to the tree-level value given in
Eq.~(\ref{eq:treelevelW}).  At this level of analysis, one therefore
only needs to compute one-loop contributions to $\hat S$ and $\hat T$,
as was done for the $N=2$ LWSM in Ref.~\cite{Chivukula:2010nw}.

\subsection{Fermion Loop Contributions}
After the tree-level contributions, the most important contributions
to the oblique parameters (indeed, the leading ones for $\hat S$ and
$\hat T$) arise from one-loop diagrams of the $t$ and $b$ quarks, as
depicted in Fig.~\ref{vacpol:fermion}.

\begin{fmffile}{vacgraphs}

\begin{figure}
\centering
\begin{eqnarray}
\nonumber
\mbox{$\Pi^f_{\hat W^+ \hat W^-} (q^2)$} = \sum_{ij} &
\parbox{120pt}{\hspace{0em}
\begin{fmfgraph*}(75,75)
\fmfleft{i}
\fmfright{o}
\fmflabel{$\hat W^+$}{i}
\fmflabel{$\hat W^+$}{o}
\fmf{photon,tension=5,label=$q \to \ $}{i,v1}
\fmf{photon,tension=5,label=$\ q \to$}{v2,o}
\fmf{fermion,left,tension=0.4,label=$t_i$}{v1,v2}
\fmf{fermion,left,tension=0.4,label=$b_j$}{v2,v1}
\fmf{phantom}{v1,v2}
\end{fmfgraph*}
} & \nonumber \\
\mbox{$\Pi^f_{\hat W^3 \hat W^3} (q^2)$} = \sum_{ij} & \left[
\parbox{280pt}{\hspace{0em}
\begin{fmfgraph*}(75,75)
\fmfleft{i}
\fmfright{o}
\fmflabel{$\hat W^3$}{i}
\fmflabel{$\hat W^3$}{o}
\fmf{photon,tension=5,label=$q \to \ $}{i,v1}
\fmf{photon,tension=5,label=$\ q \to$}{v2,o}
\fmf{fermion,left,tension=0.4,label=$t_i$}{v1,v2}
\fmf{fermion,left,tension=0.4,label=$t_j$}{v2,v1}
\fmf{phantom}{v1,v2}
\end{fmfgraph*}
\hspace{7em}
\begin{fmfgraph*}(75,75)
\fmfleft{i}
\fmfright{o}
\fmflabel{$+ \ \ \ \ \hat W^3$}{i}
\fmflabel{$\hat W^3$}{o}
\fmf{photon,tension=5,label=$q \to \ $}{i,v1}
\fmf{photon,tension=5,label=$\ q \to$}{v2,o}
\fmf{fermion,left,tension=0.4,label=$b_i$}{v1,v2}
\fmf{fermion,left,tension=0.4,label=$b_j$}{v2,v1}
\fmf{phantom}{v1,v2}
\end{fmfgraph*}
} \right] & \nonumber \\
\mbox{$\Pi^f_{\hat W^3 \hat B} (q^2)$} = \sum_{ij} & \left[
\parbox{280pt}{\hspace{0em}
\begin{fmfgraph*}(75,75)
\fmfleft{i}
\fmfright{o}
\fmflabel{$\hat W^3$}{i}
\fmflabel{$\hat B$}{o}
\fmf{photon,tension=5,label=$q \to \ $}{i,v1}
\fmf{photon,tension=5,label=$\ q \to$}{v2,o}
\fmf{fermion,left,tension=0.4,label=$t_i$}{v1,v2}
\fmf{fermion,left,tension=0.4,label=$t_j$}{v2,v1}
\fmf{phantom}{v1,v2}
\end{fmfgraph*}
\hspace{7em}
\begin{fmfgraph*}(75,75)
\fmfleft{i}
\fmfright{o}
\fmflabel{$+ \ \ \ \ \hat W^3$}{i}
\fmflabel{$\hat B$}{o}
\fmf{photon,tension=5,label=$q \to \ $}{i,v1}
\fmf{photon,tension=5,label=$\ q \to$}{v2,o}
\fmf{fermion,left,tension=0.4,label=$b_i$}{v1,v2}
\fmf{fermion,left,tension=0.4,label=$b_j$}{v2,v1}
\fmf{phantom}{v1,v2}
\end{fmfgraph*}
} \right] & \nonumber \\
\mbox{$\Pi^f_{\hat B \hat B} (q^2)$} = \sum_{ij} & \left[
\parbox{280pt}{\hspace{0em}
\begin{fmfgraph*}(75,75)
\fmfleft{i}
\fmfright{o}
\fmflabel{$\hat B$}{i}
\fmflabel{$\hat B$}{o}
\fmf{photon,tension=5,label=$q \to \ $}{i,v1}
\fmf{photon,tension=5,label=$\ q \to$}{v2,o}
\fmf{fermion,left,tension=0.4,label=$t_i$}{v1,v2}
\fmf{fermion,left,tension=0.4,label=$t_j$}{v2,v1}
\fmf{phantom}{v1,v2}
\end{fmfgraph*}
\hspace{6em}
\begin{fmfgraph*}(75,75)
\fmfleft{i}
\fmfright{o}
\fmflabel{$+ \ \ \ \ \hat B$}{i}
\fmflabel{$\hat B$}{o}
\fmf{photon,tension=5,label=$q \to \ $}{i,v1}
\fmf{photon,tension=5,label=$\ q \to$}{v2,o}
\fmf{fermion,left,tension=0.4,label=$b_i$}{v1,v2}
\fmf{fermion,left,tension=0.4,label=$b_j$}{v2,v1}
\fmf{phantom}{v1,v2}
\end{fmfgraph*}
} \right] & \nonumber
\end{eqnarray}
\caption{Fermion vacuum polarization Feynman diagrams that provide the
dominant contributions to the electroweak precision observables $\hat
S$ and $\hat T$.}
\label{vacpol:fermion}
\end{figure}

\end{fmffile}

Consider the one-loop fermionic contributions to the self-energy
connecting generic gauge bosons $\hat{A}$ and $\hat{B}$ (the latter
not to be confused with the actual $\hat{B}$ field in the Standard
Model).  To do so, we begin with mass-diagonalized fermion fields
labeled by $i,j$, and write the interaction Lagrangian:
\begin{equation}
\mathcal{L}=\bar{\Psi}^0_i \gamma^{\mu}[\hat{A}_{\mu}
(A^{L,\Psi}_{ij}P_L + A^{R,\Psi}_{ij}P_R)+\hat{B}_{\mu}
(B^{L,\Psi}_{ij}P_L+ B^{R,\Psi}_{ij}P_R) ] \Psi^0_j \, .
\end{equation}
The fermionic mass eigenstate fields $(\Psi_i^0)^T$ are defined by
combining Eqs.~(\ref{superfields}) and (\ref{masseigen}).  The
coupling matrices are the charges in mass basis, {\it e.g.},
$A^{L,\Psi}_{ij}=S^{\Psi \, \dagger}_L Q_{A,L}^\Psi \eta S^{\Psi}_L$.
Here, $Q_A^\Psi$ is the matrix of fermion charges under the gauge
group $A$, and the superscript $\Psi$ may refer to a single flavor (as
for $\gamma$, $Z^0$) or a specific flavor transition (as for $W^\pm$).
The right-handed coupling matrices are obtained by exchanging
$L\leftrightarrow R$.

In accord with the noncanonical normalization of fields inherited by
the polarization functions in Eqs.~(\ref{eq:ggprime})--(\ref{eq:GF}),
the fermionic one-loop contribution to the self-energy contains no
gauge coupling constants, and is expressed as:
\begin{eqnarray}
\lefteqn{\Pi_{AB}(q^2) = \frac{C}{8\pi^2}} & &  \nonumber \\
& \times & \displaystyle\sum_{\Psi=T,B}\displaystyle\sum_{i,j}
\eta_{ii}\eta_{jj}\left[(A^{L,\Psi}_{ij}B^{L,\Psi}_{ji}+
A^{R,\Psi}_{ij}B^{R,\Psi}_{ji})I_1(q^2)+ (A^{L,\Psi}_{ij}
B^{R,\Psi}_{ji}+A^{R,\Psi}_{ij}B^{L,\Psi}_{ji})I_2(q^2)m_im_j\right]
\, , \nonumber \\ \label{fermionself}
\end{eqnarray}
where $C$ is a color factor ($= N_c$ for quarks coupling to colorless
gauge bosons).  Defining $\Delta \equiv -q^2x(1-x)+m_i^2x+m_j^2(1-x)$
for the usual two-propagator factor, and using primes to indicate
$q^2$ derivatives and subscript 0 to indicate a function evaluated at
$q^2 = 0$ so that $\Delta_0 = m_i^2x+m_j^2(1-x)$, $\Delta'_0=-x(1-x)$,
and $\Delta''_0=0$, the integrals are defined as follows:
\begin{eqnarray}
I_1(q^2) & \equiv & \int_0^1dx~
\left( 2\Delta - \Delta_0 \right) \, \ln (\Delta/M^2) \, , \\
I_2(q^2) & \equiv & -\int_0^1dx~
\ln (\Delta/M^2) \, .
\end{eqnarray}
One then obtains the moments of the integrals relevant to the oblique
parameters:
\begin{eqnarray}
I_{10} & = & \int_0^1dx~
\Delta_0 \, \ln (\Delta_0/M^2) \, , \\
I_{20} & = & -\int_0^1dx~
\ln (\Delta_0/M^2) \, , \\
I'_{10} & = & \int_0^1dx~
\Delta'_0 [1+ 2\ln (\Delta_0/M^2)] \, , \\
I'_{20} & = & -\int_0^1dx~
\Delta'_0/\Delta_0 \, , \\
I''_{10} & = & 3 \int_0^1dx~
(\Delta'_0)^2/\Delta_0 \, , \\
I''_{20} & = & \int_0^1dx~
(\Delta'_0/\Delta_0)^2 \, .
\end{eqnarray}
The factor $M^2$ contains the parameter of the logarithmic divergence
and various subtraction constants associated with the regularization
procedure.  Of course, $M^2$ must cancel from the complete expressions
for the oblique parameters, since they are observables.  The
individual integrals are straightforward and give:
\begin{eqnarray}
I_{10} & = & -\frac{1}{4} (m_i^2 + m_j^2) + \frac{1}{2} \frac{m_i^4
\ln (m_i^2/M^2) - m_j^4 \ln (m_j^2/M^2)}{m_i^2 - m_j^2} \, , \nonumber
\\ & \to & m_i^2 \ln \frac{m_i^2}{M^2} \, , \ \ m_j \to m_i \, ; \\
I_{20} & = & 1 - \frac{m_i^2 \ln (m_i^2/M^2) - m_j^2 \ln
(m_j^2/M^2)}{m_i^2 - m_j^2} \, , \nonumber \\ & \to &
- \ln \frac{m_i^2}{M^2} \, , \ \ m_j \to m_i \, ; \\
I'_{10} & = & -\frac{1}{3} \left\{ \frac{m_i^4 (m_i^2 -
3m_j^2)}{(m_i^2 - m_j^2)^3} \ln \left( \frac{m_i^2}{M^2} \right)
\! - \frac{m_j^4 (m_j^2 - 3m_i^2)}{(m_i^2 - m_j^2)^3}
\ln \left( \frac{m_j^2}{M^2} \right) \! +
\frac{m_i^4 - 8 m_i^2 m_j^2 + m_j^4}{3(m_i^2 - m_j^2)^2} \right\} ,
\nonumber \\ & \to &
-\frac{1}{6} \left[ 1 + 2 \ln \left( \frac{m_i^2}{M^2} \right)
\right] \, , \ \ m_j \to m_i \, ; \\
I'_{20} & = & - \frac{(m_i m_j)^2}{(m_i^2 -
m_j^2)^3} \ln \left( \frac{m_i^2}{m_j^2} \right) + \frac{m_i^2 +
m_j^2}{2(m_i^2 - m_j^2)^2} \, , \nonumber \\
& \to & \frac{1}{6m_i^2} \, , \ \ m_j \to m_i \, ; \\
I''_{10} & = &
\frac{3 (m_i m_j)^4}{(m_i^2 - m_j^2)^5} \ln \left(
\frac{m_i^2}{m_j^2} \right) + \frac{(m_i^2 + m_j^2)(m_j^2 - 8m_i^2
m_j^2 + m_i^4)}{4(m_i^2-m_j^2)^4} \, , \nonumber \\
& \to & \frac{1}{10 m_i^2} \, , \ \ m_j \to m_i \, ; \\
I''_{20} & = & -\frac{2(m_i m_j)^2 (m_i^2 +
m_j^2)}{(m_i^2 - m_j^2)^5} \ln \left( \frac{m_i^2}{m_j^2} \right) +
\frac{m_i^4 + 10 m_i^2 m_j^2 + m_j^4}{3(m_i^2 - m_j^2)^4} \,
, \nonumber \\
& \to & \frac{1}{30 m_i^4} \, , \ \ m_j \to m_i \, .
\end{eqnarray}

These expressions are inserted into Eq.~(\ref{fermionself}) to produce
the full results for the fermionic one-loop contributions; however,
the $S_{L,R}$ matrices enter the couplings $A,B$ (and both $S_{L,R}$
are required [Eq.~(\ref{massxfm})] to produce the fermion mass
eigenvalues).  While analytic expansions for $S_{L,R}$ appear in the
literature~\cite{Alvarez:2008za,Krauss:2007bz}, in practice we perform
the calculations numerically and therefore do not present the full
cumbersome expressions for the oblique parameters.

\section{Constraints from the $Z b_L \bar b_L$ Coupling} \label{sec:dg}

\begin{figure}
\centering
\mbox{$\delta g_L^{b\bar b} \sim \sum_{ijk} \ \ $}
\parbox{140pt}{\hspace{0em}
\begin{fmfgraph*}(120,120)
\fmfleft{i}
\fmfright{o1,o2}
\fmflabel{$\phi^0$}{i}
\fmflabel{$b_L$}{o1}
\fmflabel{$\bar b_L$}{o2}
\fmf{dashes,tension=5,label=$p \to \ $}{i,v0}
\fmf{fermion,tension=5,label=$p$}{v1,o1}
\fmf{fermion,tension=5}{o2,v2}
\fmf{fermion,tension=2,label=$t_i$}{v0,v1}
\fmf{fermion,tension=2,label=$\bar t_j$}{v2,v0}
\fmf{dashes,tension=1,label=$h^+_k$}{v1,v2}
\end{fmfgraph*}
} \vspace{1em}
\caption{Dominant diagram contributing to the $Z b_L \bar b_L$
coupling.  $\phi^0$ is the Goldstone boson eaten by the $Z^0$, and
indices $i$,$j$,$k$ denote mass eigenstates.  The coupling is defined
in the limit $p \to 0$.}
\label{dgLdiagram}
\end{figure}
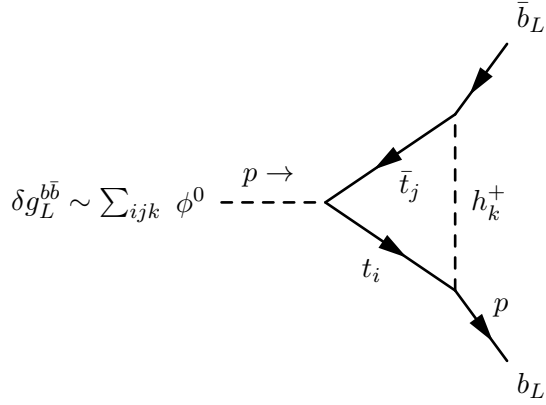

One of the more interesting direct electroweak precision observables
in terms of the tension between the experimental measurement and its
SM prediction is the $Z b_L \bar b_L$ coupling.  As noted long
ago~\cite{Barbieri:1992nz}, its leading contribution in the gaugeless
limit [{\it i.e.}, ignoring effects suppressed by $(m_{Z^0}/m_t)^2$]
is most easily obtained by computing the triangle loop diagram of
Fig.~\ref{dgLdiagram}, in which a Goldstone boson $\phi^0$ (the one
eaten by the $Z^0$) of momentum $p$ splits into a $t \bar t$ pair,
which subsequently (via exchange of a charged scalar) decays to $b_L
\bar b_L$.  The invariant amplitude for this triangle loop diagram in
the $p \to 0$ limit can be parametrized as
\begin{equation}
i {\cal M} = - \frac{2}{v} (\delta g_L^{b \bar b}) p \hspace{-0.4em} /
P_L \, .
\end{equation}
The coupling $g_L^{b \bar b}$ is derived from a combination of the $Z^0
\to b \bar b$ branching fraction $R_b$ and its forward-backward
asymmetry $A_b$; an indication of its sensitivity to small changes in
both is given in Ref.~\cite{Haber:1999zh}:
\begin{eqnarray}
\delta g_L^{b \bar b} & \equiv & g_L^{b \bar b , \, \rm exp}
- g_L^{b \bar b , \, \rm SM} \nonumber \\
& = & -1.731 \, \delta R_b -0.1502 \, \delta A_b \, ,
\label{HaberLogan}
\end{eqnarray}
where the normalization has been adjusted [{\it i.e.}, removing the
$e/(\sin \theta_W \cos \theta_W)$ coefficient] to match that used
elsewhere in this section.  Its most recent experimental value $g_L^{b
\bar b , \, \rm exp} = -0.4182(15)$ has not changed since the combined
LEP/SLD 2005 analysis~\cite{ALEPH:2005ab}.  The SM value $g_L^{b \bar
b , \, \rm SM} = -0.42114^{+45}_{-24}$ from~\cite{ALEPH:2005ab} gives
$\delta g_L^{b \bar b} = +2.94(157) \cdot 10^{-3}$, meaning that the
SM value was $\approx 2\sigma$ low, thus strongly disfavoring any new
physics contribution with $\delta g_L^{b \bar b} < 0$.  The current
Particle Data Group~\cite{Beringer:1900zz} values for $R_b^{\rm SM}$
and $A_b^{\rm SM}$, however, lead [via Eq.~(\ref{HaberLogan})] to a
somewhat relaxed bound,
\begin{equation}
\delta g_L^{b \bar b} = +2.69(157) \cdot 10^{-3} \, ,
\label{dgLbound}
\end{equation}
which we use in our analysis.

\begin{fmffile}{yukgraph}

\end{fmffile}

The effect of $N=2$ LWSM states on $\delta g_L^{b\bar b}$ has been
considered twice in the literature.  The central result of
Ref.~\cite{Carone:2009nu} is that current precision bounds allow LW
Higgs partner masses to be significantly lighter than other LW states.
Therefore, \cite{Carone:2009nu} effectively compute $\delta g_L^{b\bar
b}$ including only a LW Higgs partner in the triangle loop diagram,
giving (in our normalization):
\begin{equation} \label{CLgL}
\delta g_L^{b \bar b} = -\frac{m_t^2}{16\pi^2 v^2} \left[
\frac{R}{R-1} - \frac{R \ln R}{(R-1)^2} \right] \, ,
\end{equation}
where $R = (m_t/m_{h_2})^2$, so that $\delta g_L^{b \bar b} < 0$.
$\delta g_R^{b \bar b}$ in the LWSM is driven by $m_b$ and hence is
numerically much smaller.  Since $\delta g_L^{b \bar b}$ and $\delta
R_b$ are anti-correlated [Eq.~(\ref{HaberLogan})], and since $\delta
R_b$ is positive~\cite{ALEPH:2005ab,Beringer:1900zz},
Ref.~\cite{Carone:2009nu} then states that the LW Higgs partner
contribution acts in the direction of reconciling the discrepancy, and
concludes that $\delta g_L^{b \bar b}$ analysis gives no meaningful
bound on the LW scalar mass.  However, Eq.~(\ref{HaberLogan}) shows
that $\delta g_L^{b \bar b}$ also depends strongly upon $\delta A_b$,
and the combined effect is to create the situation described above, in
which new physics $\delta g_L^{b \bar b} < 0$ contributions are
actually more difficult to accommodate.  We take this additional
effect into account in our analysis.

On the other hand, Ref.~\cite{Chivukula:2010nw} uses the full $\delta
g_L^{b \bar b}$ bound from~\cite{ALEPH:2005ab,Beringer:1900zz}
described above, but includes only LW $t$-quark partners in the
triangle diagram, thus producing the result
\begin{equation}
\delta g_L^{b \bar b} = -\frac{m_t^4}{32\pi^2 v^2 M_q^2}
\left[ 5 \ln \frac{M_q^2}{m_t^2} - \frac{49}{6} \right] \, ,
\label{CFFSgL}
\end{equation}
at leading orders in $m_t^2 / M_q^2$.  The result
of~\cite{Chivukula:2010nw} obtained from this observable is the most
stringent in their entire analysis, giving a lower bound of $M_q \agt
4$~TeV\@.  However, the LW correction (\ref{CFFSgL}) is a very shallow
function of $M_q$ (see their Fig.~8), and the small change in the SM
value of $g_L^{b \bar b}$ described above is alone enough to push the
bound back to about $M_q \agt 1.2$~TeV\@.  Obviously, the contribution
from the LW Higgs partner must also be included in a global analysis,
and since it is also negative (and indeed, turns out to be comparable
in magnitude to the LW $t$ contribution), all of the mass lower bounds
in such a circumstance would be higher, but these multiple
considerations should serve to illustrate that room exists in mass
parameter space to accommodate interesting LWSM possibilities even in
the $N=2$ case.
%
%

Here, we examine the $N=3$ LWSM contribution to $\delta g_L^{b \bar
b}$; since the $N=2$ effect was computed in
Ref.~\cite{Chivukula:2010nw}, we closely follow the notation
introduced there.  The Yukawa Lagrangian
\begin{eqnarray} \label{LYuk}
{\cal L}_{\rm Yuk} = -i y_t \sum_{i,j} \left\{ \frac{1}{\sqrt{2}}
\hat \phi^0 \left[ \alpha_{ij} \bar t_i P_R t_j - \alpha_{ji}
\bar t_i P_L t_j \right] + \beta_{ij} \left[ \hat \phi^-
\bar b_i P_R t_j - \hat \phi^+ \bar t_j P_L b_i \right] \right\} \,
\end{eqnarray}
has couplings $\alpha$ and $\beta$ closely related to the ones
appearing in the mass matrix (\ref{massmatrix}) with the Dirac mass
parameters excluded.  Specifically,
\begin{eqnarray}
\alpha \equiv (S_L^t)^\dagger \alpha_0 S_R^t \, , \nonumber \\
\beta  \equiv (S_L^b)^\dagger \beta_0  S_R^t \, ,
\end{eqnarray}
where, for the example of the $N=3$ case,
\begin{equation}
\alpha^{N=3}_0 = \beta^{N=3}_0 \equiv \left( \begin{array}{ccccc}
1 & - \cosh \phi_q & 0 & \sinh \phi_q & 0 \\
-\cosh \phi_t & \cosh \phi_q \, \cosh \phi_t & 0 &
-\sinh \phi_q \, \cosh \phi_t & 0 \\
0 & 0 & 0 & 0 & 0 \\
\sinh \phi_t & -\cosh \phi_q \, \sinh \phi_t & 0 & \sinh
\phi_q \, \sinh \phi_t & 0 \\
0 & 0 & 0 & 0 & 0
\end{array} \right) \, .
\end{equation}
The most important distinction between the expressions here and those
in Ref.~\cite{Chivukula:2010nw} is actually not the addition of the $N
= 3$ fermion partners, but rather the presence of the entire HD scalar
fields $\hat \phi^0$, $\hat \phi^\pm$ whose SM content is the set of
Goldstone bosons, and that enter with the relative weights as in
Eq.~(\ref{scalarweights}).  As indicated in
Eq.~(\ref{Higgses})--(\ref{scalarmasses}), the LW partners to these
fields are physical, massive states that must be included in the
calculation of $\delta g_L^{b \bar b}$ but were omitted in
Ref.~\cite{Chivukula:2010nw}.

The basic result of the $\delta g_L^{b \bar b}$ calculation in
Ref.~\cite{Chivukula:2010nw} is that the LW $t$-quark partners in the
loop tend to slightly exacerbate the tension with the measured value,
thus forcing an even more stringent lower bound on the LW quark mass
(4~TeV) than that obtained from $\hat T$.  As pointed out in
Ref.~\cite{Carone:2009nu}, however, the heavy $h_2^\pm$ can be much
lighter ($\agt 500$~GeV) and still satisfy all precision constraints.
Noting first from Eq.~(\ref{scalarmasses}) that the charged scalar
masses do not mix, and recalling that the virtual scalar in the
$\delta g_L^{b \bar b}$ diagram is charged, the extra signs in the
$h_{2,3}^\pm$ propagators can be used to oppose the contribution from
the original diagram with a virtual $\phi^\pm$, thus relieving much of
the additional tension in $\delta g_L^{b \bar b}$.  The full
expression reads
\begin{eqnarray}
\delta g^{b \bar b}_L & = & \frac{1}{16\pi^2} \cdot \frac{y_t^3v}
{2\sqrt{2}} \left\{ \sum_i \eta_k \beta_{0i}^2 \alpha_{ii}
\frac{m_{t_i}}{m_{t_i}^2 - m_{h_k}^2} \left[ 1 -
\frac{m_{h_k}^2}{m_{t_i}^2 - m_{h_k}^2} \ln \left(
\frac{m_{t_i}^2}{m_{h_k}^2} \right) \right] \right. \nonumber \\
& + & \sum_{i \neq j; \, k} (-1)^{i+j} \eta_k \beta_{0i}
\beta_{0j} \alpha_{ji} m_{t_j} \left[
\frac{-1}{m_{t_i}^2 - m_{t_j}^2} \cdot \frac 1 2 \left(
\frac{m_{t_i}^2}{m_{t_i}^2 - m_{h_k}^2} + \frac{m_{t_j}^2}{m_{t_j}^2 -
m_{h_k}^2} \right) \right. \nonumber \\ & + &
\frac{m_{t_i}^2}{2(m_{t_i}^2 - m_{t_j}^2)^2} \left( \frac{2 m_{t_i}^2
- m_{t_j}^2}{m_{t_i}^2 - m_{h_k}^2} + \frac{m_{t_j}^2}{m_{t_j}^2 -
m_{h_k}^2} \right) \ln
\left( \frac{m_{t_i}^2}{m_{t_j}^2} \right) \nonumber \\
& - & \frac{m_{h_k}^2}{2(m_{t_i}^2-m_{h_k}^2)(m_{t_j}^2-m_{h_k}^2)}
\left[ \frac{2m_{t_i}^2 - m_{h_k}^2}{m_{t_i}^2 - m_{h_k}^2} \ln \left(
  \frac{m_{t_j}^2}{m_{h_k}^2} \right) - \frac{m_{h_k}^2}{m_{t_j}^2 -
  m_{h_k}^2} \ln \left( \frac{m_{t_i}^2}{m_{h_k}^2} \right) \right]
\nonumber
\\ & - & \left. \left. \frac{m_{h_k}^2}{2(m_{t_i}^2-m_{t_j}^2)}
\ln \left( \frac{m_{t_i}^2}{m_{t_j}^2} \right) \left(
  \frac{m_{t_i}^2}{(m_{t_i}^2 - m_{h_k}^2)^2} -
\frac{m_{t_j}^2}{(m_{t_j}^2 - m_{h_k}^2)^2}\right) \right] \right\}
\, .
\end{eqnarray}
The coefficients $\eta_k$ here are ones that appear in
Eq.~(\ref{scalarweights}).  This expression reduces, in the limits
$m_{h_1} \to 0$ and $m_{h_{2,3}} \to \infty$, to Eq.~(A6) of
Ref.~\cite{Chivukula:2010nw} [which, in turn, reduces to
Eq.~(\ref{CFFSgL}) in the further limit $m_t \ll m_{t_{2,3}}$].
Alternately, it reduces in the limit $m_{t_{2,3}}$, $m_{h_3} \to
\infty$ to Eq.~(\ref{CLgL}), as was used in Ref.~\cite{Carone:2009nu}.

\section{Analysis} \label{sec:analysis}

We use the definitions of the post-LEP oblique parameters in
Eqs.~(\ref{eq:treelevelShat})--(\ref{eq:treelevelY}).  As discussed
above, the tree-level expressions for $W$ and $Y$ are sufficient for
our analysis (and provide the most useful bounds on electroweak gauge
boson partner masses), while the leading contributions to $\hat S$ and
$\hat T$ arise from one-loop fermion effects.  Since the sums in
Eq.~(\ref{fermionself}) include the SM quarks, their effects must be
subtracted from the full result, giving $\hat S_{\rm new} \equiv \hat
S - \hat S_{\rm SM}$ and $\hat T_{\rm new} \equiv \hat T - \hat T_{\rm
SM}$.  In our subsequent discussion, $\hat S$, $\hat T$ are understood
to mean $\hat S_{\rm new}$, $\hat T_{\rm new}$, respectively.  As a
benchmark for the magnitude of new physics effects, one finds $\hat
S_{\rm SM} = -1.98 \cdot 10^{-3}$, $\hat T_{\rm SM} = +9.25 \cdot
10^{-3}$.

As seen in Ref.~\cite{Barbieri:2004qk}, the measured values of the
parameters $\hat S$, $\hat T$, $W$, and $Y$ are all of order
$10^{-3}$, and they are correlated.  However, for simplicity we use
the values listed in Table~4 of~\cite{Barbieri:2004qk} with 2$\sigma$
uncertainties:
%
\begin{eqnarray}
10^3 \, \hat S & = &  0.0 \pm 2.6 \, , \\
10^3 \, \hat T & = &  0.1 \pm 1.8 \, , \\
10^3 \, W      & = & -0.4 \pm 1.6 \, , \label{Wbound} \\
10^3 \, Y      & = &  0.1 \pm 2.4 \, .
\end{eqnarray}
To this list we add the bound on $\delta g_L^{b\bar b}$ in
Eq.~(\ref{dgLbound}), which serves to constrain both LW fermion masses
and scalar masses, as discussed in the previous section.

\begin{figure}
  \centering
  \includegraphics[width=80mm]{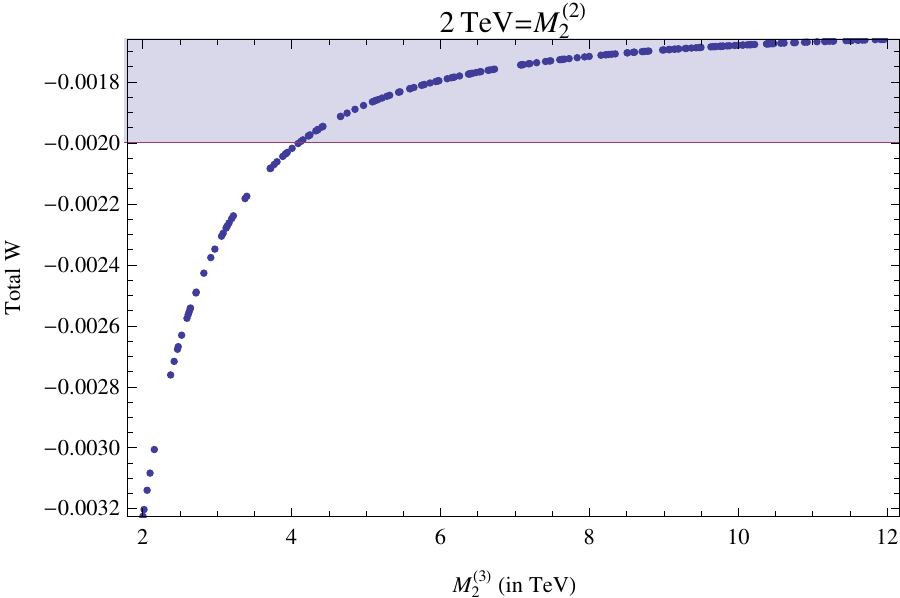}
\includegraphics[width=80mm]{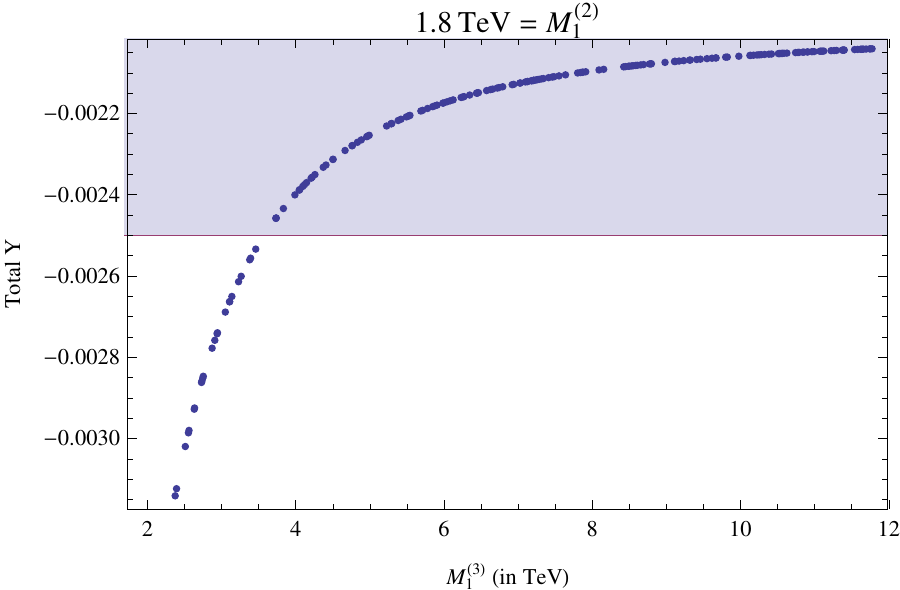}
\caption{Bounds on LW gauge boson mass partners from the oblique
  parameters $W$ and $Y$.  The shaded area (blue online) is
  experimentally allowed at 2$\sigma$.}
  \label{WYplots}
\end{figure}

First note that the $N=2$ and $N=3$ gauge boson masses contribute at
tree level in Eqs.~(\ref{eq:treelevelW})--(\ref{eq:treelevelY})
additively, and therefore the bounds that hold for the $N=2$ theory
({\it e.g.}, $M_1^{(2)} = M_2^{(2)} \geq 2.4$~TeV according to
Ref.~\cite{Chivukula:2010nw}) are tightened by the addition of $N=3$
partners.  In Fig.~\ref{WYplots} one sees that taking $M_2^{(2)} =
2$~TeV requires $M_2^{(3)} \agt 4$~TeV, the latter likely outside the
discovery range of the current LHC\@.  In particular, the discovery
scenario described in Ref.~\cite{Lebed:2012zv} of $M_2^{(2)} =
2.0$~TeV, $M_2^{(3)} = 2.5$~TeV is unlikely unless the bounds on $W$
are not as stringent as given in Eq.~(\ref{Wbound}).  Likewise, for
$Y$, Fig.~\ref{WYplots} indicates $M_1^{(2)} = 1.8$~TeV is possible
for $M_1^{(3)} \agt 3.5$~TeV\@.  If, however, the $N=2$ and $N=3$
masses are quasi-degenerate, universal values $\agt 2.5$~TeV remain
possible.

The constraints from $\hat S$ are much less restrictive.  Unlike in
other BSM scenarios where the addition of extra chiral fermions create
insurmountable tension with the measured value of $\hat S$, the extra
fermions in the LWSM are all vectorlike, and contribute to $\hat S$
only through diagonalization with the chiral fermion mass parameters
arising through Yukawa couplings.  Assuming for simplicity the
degenerate case $M_{q2} = M_{t2} = M_{b2}$ studied
in~\cite{Chivukula:2010nw} and extending to $M_{q3} = M_{t3} =
M_{b3}$, one finds no meaningful constraint on the fermion mass
parameters $M_{q2}$ or $M_{q3}$.

\begin{figure}
  \centering
  \includegraphics[width=80mm]{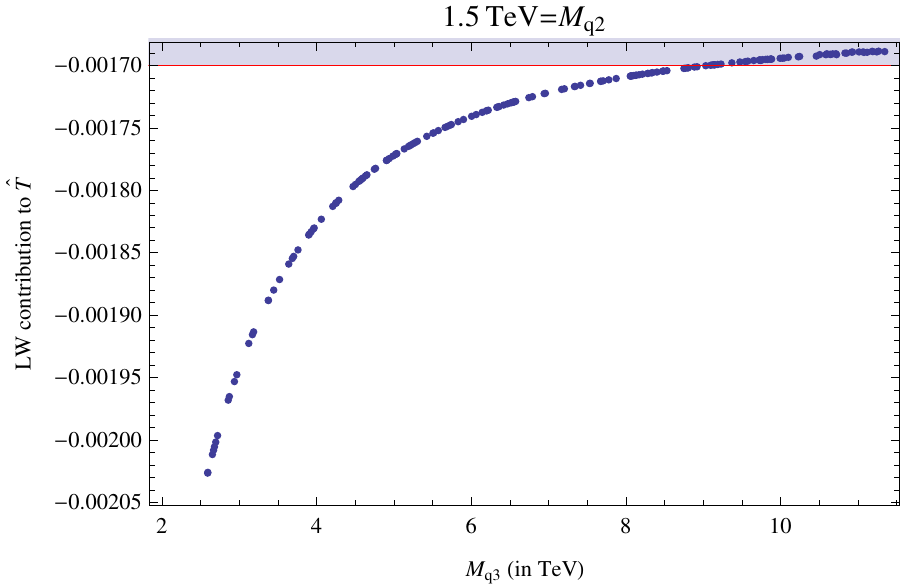}
  \includegraphics[width=80mm]{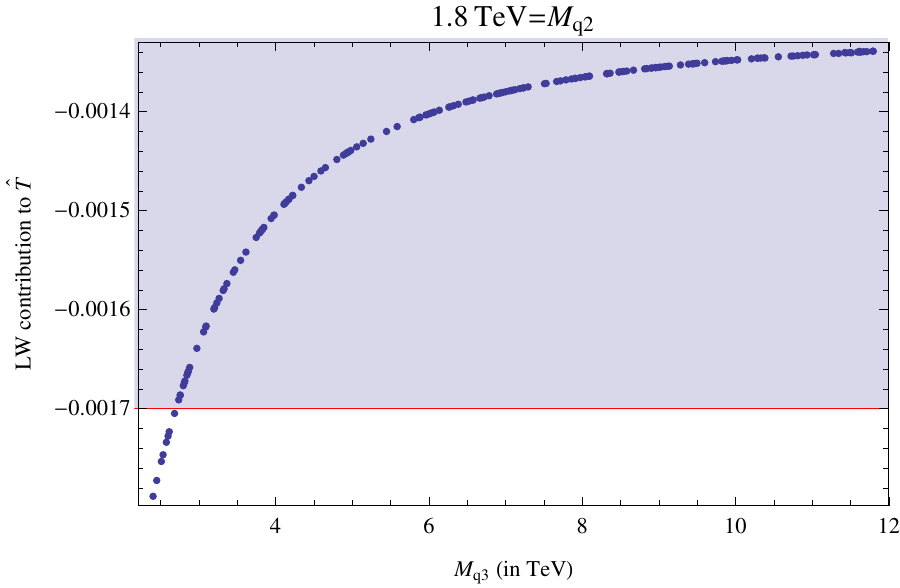}
  \caption{Bounds on the oblique parameter $\hat T$ in two scenarios,
    $M_{q2} = 1.5$~TeV and 1.8~TeV\@.  The shaded area (blue online)
    is experimentally allowed at 2$\sigma$.}
  \label{Thatplots}
\end{figure}

The bounds from $\hat T$ are much more interesting; they were found
in~\cite{Chivukula:2010nw} (Fig.~5) to require $M_{q2} \geq 1.5$~TeV
in order for $\hat T$ to lie no more than $2\sigma$ below its measured
central value, and provide one of the strongest constraints on LW
quark partner masses.  At the inception of this work, it was believed
that the opposite signs of the $N=2$ and $N=3$ LW quark propagators
would allow for a near-complete cancellation of their loop effects,
essentially removing the $\hat T$ constraint as a significant bound on
the quark partners if their masses were sufficiently close.  However,
the detailed result in fact requires much greater care in its
analysis: While the $N=2$ and $N=3$ loops do indeed cancel to a large
extent, the propagating fermions in the loops are the mass
eigenstates.  The act of mass diagonalization not only shifts mass
eigenvalues of the heavy states slightly away from $M_{q2}$ and
$M_{q3}$, but also modifies the strength of the contribution of the
$N=1$ (SM) quarks to $\hat T$.  The effect of this shift is pronounced
due to the large size of the SM $t$ Yukawa coupling; it actually
serves to push the full value of $\hat T$ slightly further from its
measured central value, thus forcing an allowable $N=2$ LW mass
$M_{q2}$ to be slightly larger than before the addition of the $N=3$
state.  However, the effect is not extreme; from Fig.~\ref{Thatplots},
one sees that $M_{q2} = 1.5$~TeV remains viable for $M_{q3} \agt
9$~TeV, while increasing $M_{q2}$ only slightly, to 1.8~TeV, allows
$M_{q3}$ to be $\alt 2.8$~TeV\@.  The transition between extremely
strong and extremely weak $M_{q3}$ bounds occurs in a very narrow
window of $M_{q2}$ values.

\begin{figure}
  \centering
  \includegraphics[width=80mm]{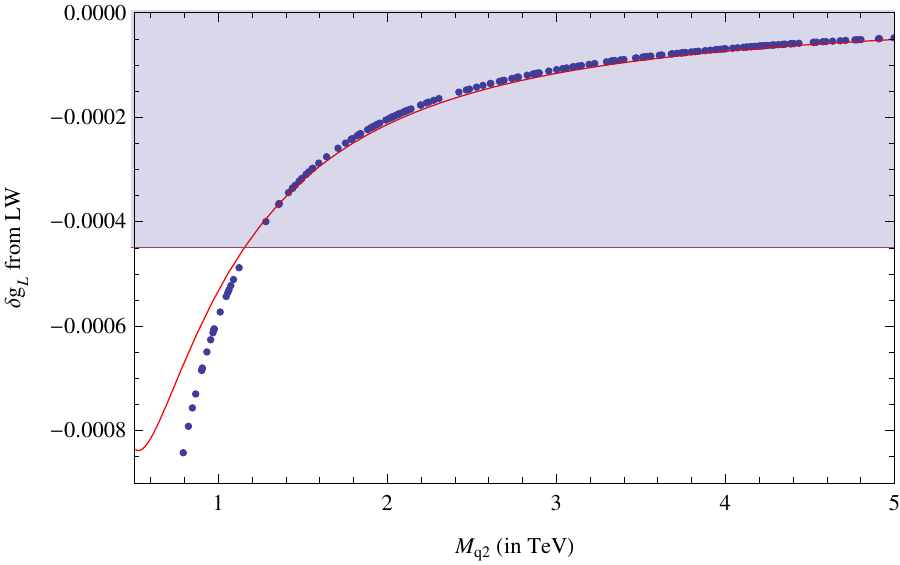}
  \includegraphics[width=80mm]{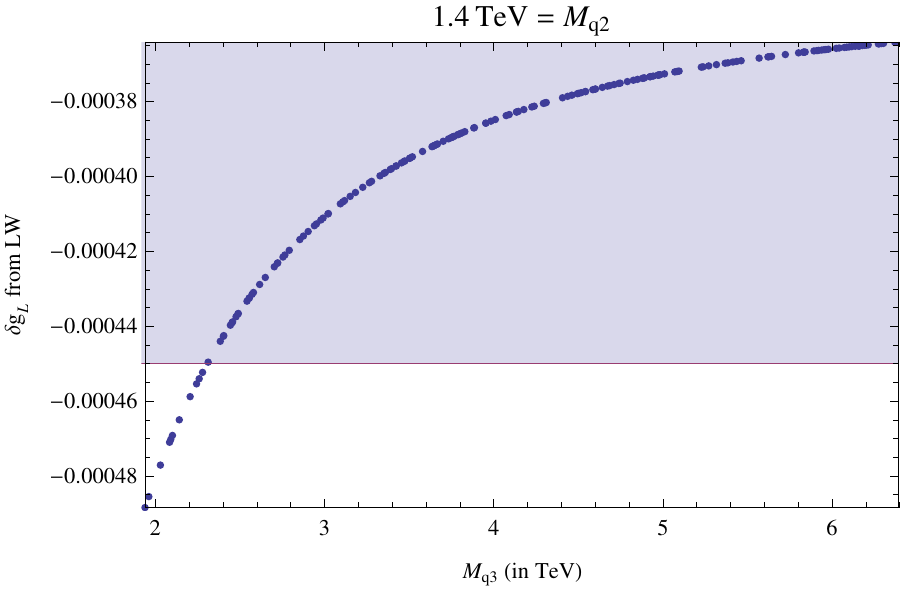}
  \caption{Bounds on $\delta g_L^{b \bar b}$ with LW $t$ quark
    partners.  The first inset presents an updated calculation in the
    $N=2$ theory, and the red line is the leading-order result
    Eq.~(\ref{CFFSgL}).  The second inset presents an $N=3$ calculation
    in which $M_{q2}$ is fixed and $M_{q3}$ is allowed to vary.  The
    shaded area (blue online) is experimentally allowed at 2$\sigma$.}
  \label{dgLquarks}
\end{figure}

\begin{figure}
  \centering
  \includegraphics[width=80mm]{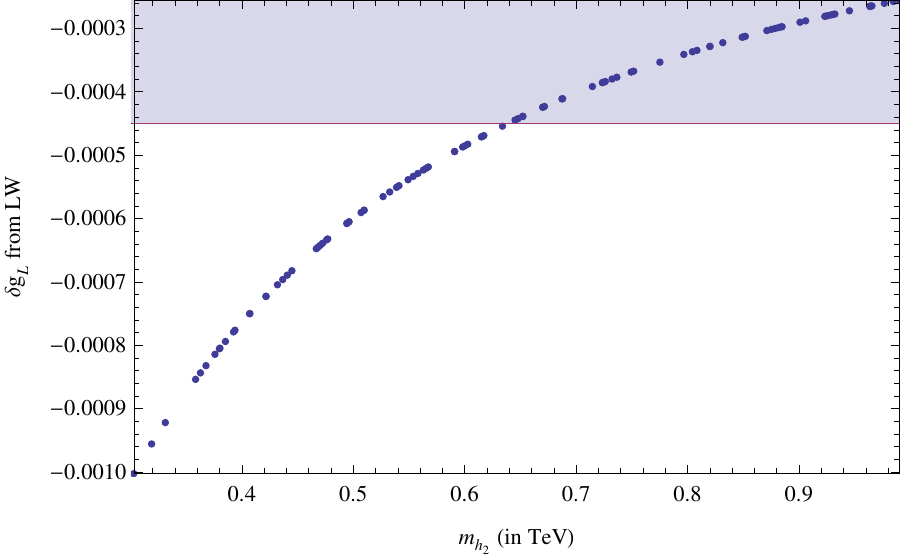}
  \includegraphics[width=80mm]{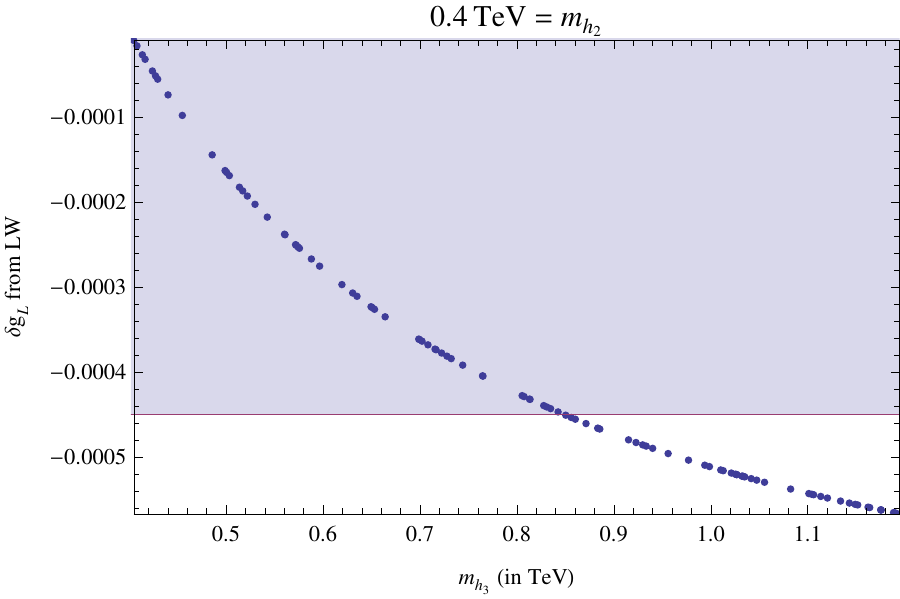}
  \caption{Bounds on $\delta g_L^{b \bar b}$ with one (first inset,
    $N=2$) and two (second inset, $N=3$) charged scalar LW partners,
    of masses $m_{h_{2,3}}$.  The shaded area (blue online) is
    experimentally allowed at 2$\sigma$.}
  \label{N=3Higgs}
\end{figure}

\begin{figure}
  \centering
  \includegraphics[width=80mm]{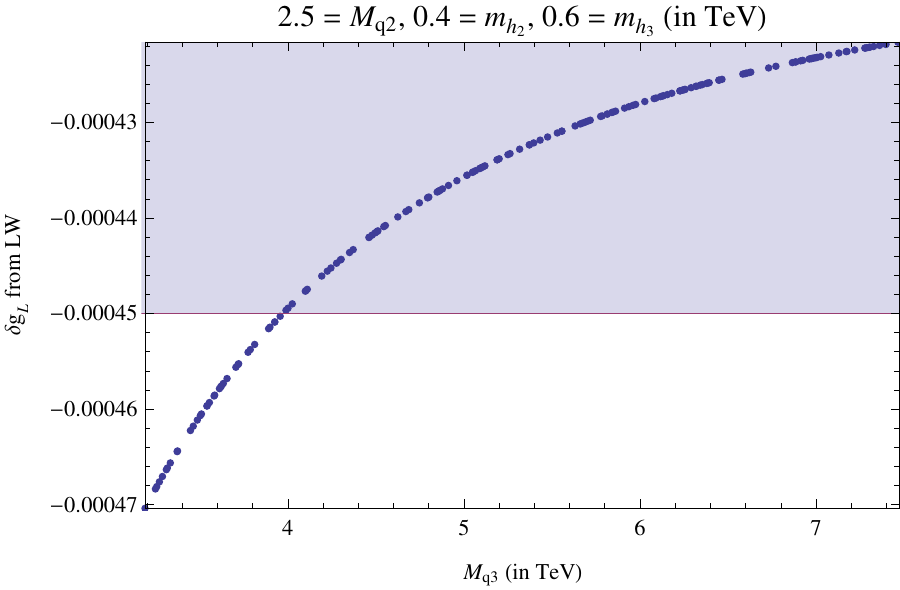}
  \caption{Bounds on $\delta g_L^{b \bar b}$ in the $N=3$ theory with
  both LW quark and charged scalar partners.  The shaded area (blue
  online) is experimentally allowed at 2$\sigma$.}
  \label{combined}
\end{figure}

Finally, consider constraints from $\delta g_L^{b \bar b}$, which in
Ref.~\cite{Chivukula:2010nw} provide the most stringent bounds on the
quark partner masses, $M_{q2} \agt 4$~TeV\@.  However, as noted in the
previous section, the bottom of the $2\sigma$-allowed region has since
moved slightly downward.  Since $\delta g_L^{b \bar b}$ is a very
shallow function of $M_{q2}$, this small change dramatically alters
the bound to $M_{q2} \agt 1.2$~TeV, as seen in the first inset of
Fig.~\ref{dgLquarks}.  The $N=3$ theory is used in the second inset of
Fig.~\ref{dgLquarks}, where one sees that raising $M_{q2}$ only
slightly (to 1.4~TeV) allows $M_{q3} \agt 2.3$~TeV\@.  On the other
hand, if the LW quark masses are assumed sufficiently large to
decouple, $\delta g_L^{b \bar b}$ provides a lower bound on the $N=2$
LW scalar of $m_{h_2} \agt 640$~GeV (first inset of
Fig.~\ref{N=3Higgs}), as would have been found in a more complete
calculation (including not only $R_b$ but also $A_b$ bounds) by
Ref.~\cite{Carone:2009nu}.  Since mass diagonalization does not mix
the charged scalar parameters, including the $N=3$ LW state leads to a
dramatic cancellation: For example, in the second inset of
Fig.~\ref{N=3Higgs} one sees that $m_{h_2} = 400$~GeV, $m_{h_3} \alt
850$~GeV satisfy the $\delta g_L^{b \bar b}$ constraint.  In
retrospect, the bounds on charged scalar masses in the $N=2$ theory
obtained by Ref.~\cite{Carone:2009nu} from $B \bar B$ mixing and $b
\to s \gamma$ now lead to weaker constraints ($m_{h_2} > 463$~GeV)
than that from $\delta g_L^{b \bar b}$, and the former bounds moreover
would also likely be significantly softened by the addition of an
$N=3$ charged scalar due to the cancellations described above.  When
both LW quarks and charged scalars are included, the bounds again
become more constrained, but many interesting scenarios remain
possible; for example, Fig.~\ref{combined} shows that the combined set
$M_{q2} = 2.5$~TeV, $M_{q3} = 4$~TeV, $m_{h_{2}} = 400$~GeV,
$m_{h_{3}} = 600$~GeV satisfies the $\delta g_L^{b \bar b}$
constraint.

\section{Discussion and Conclusions} \label{sec:concl}

The Lee-Wick approach to extending the Standard Model provides a
variety of interesting effects that can be tested experimentally.
Since the couplings of the new particles equal those of the SM fields
and only their masses remain as free parameters, one can obtain bounds
on these masses from electroweak precision constraints.  For such
particles for which the masses are $\alt 3$~TeV, one can even hope to
directly produce the particles at the current incarnation of the
LHC\@.  On the other hand, the LWSM was originally motivated by its
potential to provide an alternate resolution to the hierarchy problem,
which ideally requires fields with masses in the several hundred GeV
range.  In our calculations, we find that only the scalar partners to
the Higgs can be so light, and therefore the LWSM does not offer an
especially natural resolution of the hierarchy, although by
construction all quadratic divergences in loop diagrams cancel.

Nevertheless, we find that the imposition of precision constraints on
the $N=3$ LWSM still allows masses for LW partner states to lie in
large swathes of the parameter space directly accessible at the LHC,
providing phenomenological significance to the LWSM\@. In particular,
we have found that the post-LEP oblique parameters $W$ and $Y$ require
the $N=2$ partners of the $W$ and $B$ to be $\agt 2.0$ and $1.8$~TeV,
respectively, and the $N=3$ partners to be substantially heavier, or,
by the same bound, they could be quasi-degenerate and all $\agt
2.5$~TeV\@.  The LW quark masses are constrained by custodial isospin
($\hat T$) and the $Zb\bar b$ coupling $g_L^{b \bar b}$ to be at least
1.5~TeV; one of the most interesting results of this work was the
discovery that, as expected, the $N=3$ quarks loops do cancel against
the $N=2$ loops, but this cancellation is largely nullified by the
effects arising from the diagonalization of quark masses amongst the
SM quarks and its LW partners.  Even so, LW quark masses in the range
$M_{q2} \agt 1.8$~TeV remain viable if the $N=3$ partner is somewhat
heavier ($\agt 2.8$~TeV).  The least constrained masses, like in the
original SM, appear to be in the scalar sector.  From the $Zb \bar b$
coupling alone, values in the few hundred GeV range remain viable in
the $N=3$ theory due to the presence of a more complete cancellation
between the $N=2$ and $N=3$ states, although a full analysis including
$b \to s \gamma$ and $B \bar B$ mixing should be undertaken to obtain
global constraints.

In summary, the LWSM is alive and well, particularly its $N=3$
variant.  Some of the gauge boson and fermion partners may be
difficult to discern directly at the LHC, but the potential for direct
discovery remains.  The scalar sector, whose exploration is arguably
the central business of the LHC, is the least constrained and
therefore the most interesting from the immediate phenomenological
point of view.

\begin{acknowledgments}
The authors gratefully acknowledge helpful discussions with C.~Carone
and S.~Chivukula.  This work was supported in part by the National
Science Foundation under Grant Nos.\ PHY-0757394 and PHY-1068286.
\end{acknowledgments}

\end{document}